\documentclass[sigconf]{acmart}

\usepackage{xcolor}
\AtBeginDocument{%
  \providecommand\BibTeX{{%
    \normalfont B\kern-0.5em{\scshape i\kern-0.25em b}\kern-0.8em\TeX}}}

\copyrightyear{2026}
\acmYear{2026}
\setcopyright{cc}
\setcctype{by}
\acmConference[CHI EA '26]{Extended Abstracts of the 2026 CHI Conference on Human Factors in Computing Systems}{April 13--17, 2026}{Barcelona, Spain}
\acmBooktitle{Extended Abstracts of the 2026 CHI Conference on Human Factors in Computing Systems (CHI EA '26), April 13--17, 2026, Barcelona, Spain}
\acmDOI{10.1145/3772363.3798591}
\acmISBN{979-8-4007-2281-3/2026/04}

\ccsdesc[500]{Human-centered computing~HCI theory, concepts and models}
\ccsdesc[500]{Human-centered computing~Natural language interfaces}
\ccsdesc[300]{Human-centered computing~Interaction design theory, concepts and paradigms}

\begin{document}

\newcommand\todocite{\textcolor{red}{\textbf{[cite]}}}

\newcommand{\jnote}[1]
{{\color{black}{#1}}}

\newcommand{\snote}[1]
{{\color{blue}{#1}}}


\title[Beyond Anthropomorphism]{Beyond Anthropomorphism: \\a Spectrum of Interface Metaphors for LLMs}





\author{Jianna So}
\orcid{0009-0001-9692-1781}
\affiliation{%
  \department{}
  \institution{Harvard University}
  \city{Cambridge}
  \state{Massachusetts}
  \country{USA}
}
\email{jiannaso@g.harvard.edu}

\author{Connie Cheng}
\orcid{0009-0009-8819-7410}
\affiliation{%
  \department{}
  \institution{Harvard University}
  \city{Cambridge}
  \state{Massachusetts}
  \country{USA}
}
\email{connie_cheng@gsd.harvard.edu}

\author{Sonia Krishna Murthy}
\orcid{0009-0001-7344-4636}
\affiliation{%
  \department{}
  \institution{Harvard University}
  \city{Cambridge}
  \state{Massachusetts}
  \country{USA}
}
\email{soniamurthy@g.harvard.edu}

\renewcommand{\shortauthors}{So et al.}

\begin{abstract}
Anthropomorphizing conversational technology is a natural human tendency. Today, the anthropomorphic metaphor is overly reinforced across intelligent tools. Large Language Models (LLMs) are particularly anthropomorphized through interface design. While metaphors are inherently partial, anthropomorphic interfaces highlight similarities between LLMs and humans, but mask crucial differences. As a result, the metaphor is often taken literally; users treat LLMs as if they are truly human. With few safeguards in place, this extreme anthropomorphism drives users to delusion and harm. Users also experience dissonance between the ethics of using LLMs, their growing ubiquity, and limited interface alternatives. We propose repositioning anthropomorphism as a design variable, developing opposing extremes as a theoretical framework for how interface metaphors shape and can disrupt the default metaphor. We introduce a spectrum of metaphors from transparency-driven “anti-anthropomorphism” to uncanny “hyper-anthropomorphism”. These metaphors introduce materiality to interface metaphors, exposing LLMs as sociotechnical systems shaped by human labor, infrastructure, and data. This spectrum shifts interface design away from optimizing usability and toward encouraging critical engagement.

\end{abstract}

\keywords{LLM, anthropomorphism, metaphor, ethics, critical theory}

\begin{teaserfigure}
    \centering
    \includegraphics[width=\linewidth]{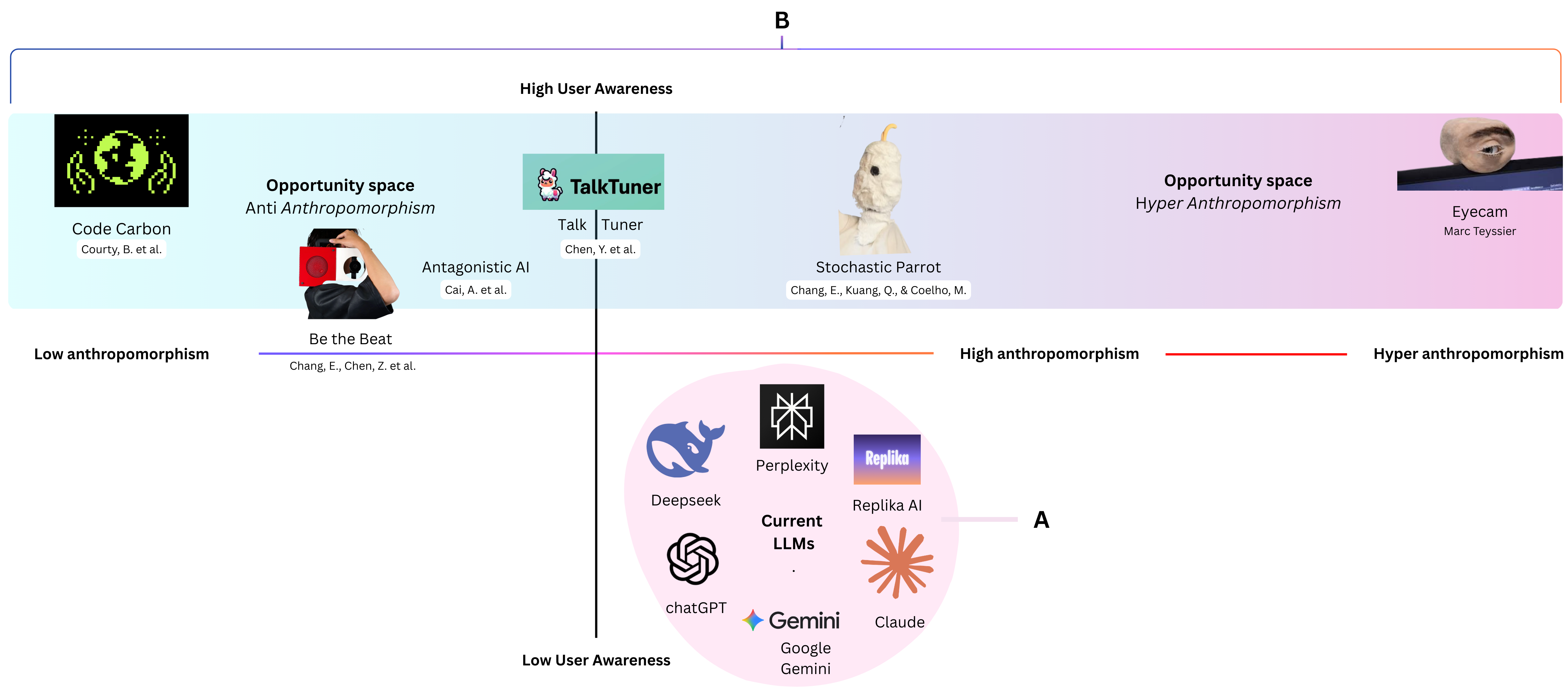}
    \caption{A diagram of existing interfaces sorted along two axes: low- to high- to hyper- anthropomorphism, and low to high user awareness of the LLM as a sociotechnical system. While current anthropomorphic interfaces  (A) result in low user awareness, we look towards an expanded spectrum of metaphors (B) that increases user awareness.}
    \label{fig:diagram}
    \Description{fill in}
\end{teaserfigure}

\maketitle

\thispagestyle{empty}

\section{Introduction}
\begin{quote}
\textit{\\"There is a universal tendency among mankind to conceive all beings like themselves [...] We find human faces in the moon, armies in the clouds."} \\

--- David Hume, \textit{The Natural History of Religion} (1757)~\cite{bennett_natural_1757}

\end{quote}
Anthropomorphism---the ascription of human qualities to non-human entities---is a natural human tendency. In the field of Human-Computer Interaction (HCI), anthropomorphism has served as a powerful design metaphor, used to render technologies more approachable and socially intelligible by framing them as human-like. As early as 1945, Vannevar Bush envisioned the Memex mirroring human thought \cite{bush_as_1945}. In the 1970's, Weizenbaum's ELIZA, a text-interactive machine therapist, suggested that minimal linguistic cues might create false perceptions of human-like intelligence and emotion \cite{weizenbaum_computer_1976}. In the 1990's, Clifford Nass identified that people treat computers as social actors \cite{nass_computers_1994}. Today, we continue to imbue tools with our likeness. Large Language Models (LLMs) that can converse using intelligent, natural language (e.g., ChatGPT, Gemini, Deepseek) are the latest and most powerful manifestations of the anthropomorphic metaphor in technology. 

In this design provocation, we explore how the anthropomorphic metaphor is overly reinforced in LLMs through interface design. Metaphors in design are powerful tools: successful metaphors can create clear and useful mental models, especially for new and complex concepts, such as intelligent technology \cite{staggers_mental_1993}. 
However, metaphors are inherently partial, meant to capture both similarities \emph{and} differences \cite{lakoff_metaphors_2003}; in contrast, the anthropomorphic metaphor hides crucial differences between humans and LLMs. Cues that build trust early on snowball into false beliefs about LLM personhood during vulnerable moments. For example, \jnote{LLMs are not only \textit{like} relationship therapists, they \textit{are} relationship therapists. For many, they have become romantic partners themselves. In extreme cases, this delusion pushes users to harm, such as documented cases of infidelity \cite{noauthor_ai_nodate}, psychosis \cite{valentino_how_2026}, and suicide \cite{noauthor_mothers_2025} related to LLM use.} 

Anthropomorphism is reinforced as the dominant interface metaphor through two main pathways: language-as-interface, and non-language interface elements. \jnote{While past research and interventions have focused on language itself to identify and combat anthropomorphism \cite{cheng_dehumanizing_2025}, we explore the latter to focus on the impact and potential of LLMs' interface design. Beyond language, LLMs are crucially anthropomorphized through chat-based interfaces, which mimic messaging and social media platforms, that include turn taking and artificial thinking delays}. In terms of the interface, we chat with LLMs the same way we text our friends.  In media theory, it is commonly held that "the medium is the message" \cite{mcluhan_understanding_1964}: the medium of media delivers distinct meaning in addition to the media itself. \textbf{In LLMs, the metaphor has become the message}: the metaphor is the medium, and LLM interfaces instantiate the metaphor. Therefore, it is crucial to study the fundamental effects of the anthropomorphic metaphor through interface design. 

In this design provocation, we argue that the default anthropomorphic metaphor in LLMs can be challenged through metaphors that emphasize differences between LLMs and humans. Through design speculation \cite{dunne2013speculative}, we propose a way forward through a spectrum of metaphors from transparent \textbf{"anti-anthropomorphism"} that exposes the reality of a system to intentionally uncanny \textbf{"hyper-anthropomorphism"} that heightens discomfort. These metaphors introduce materiality to precisely expose LLMs as sociotechnical systems \cite{dhole_large_2023} with human influence: LLMs are shaped by human labor, infrastructure, and data, and also reshape users’ long-term cognitive, emotional, and social capacities. Implementing such metaphors may complicate user mental models and introduce friction, which is typically treated as undesirable in interface and product design. However, users are \textit{already} experiencing friction. Public discourse increasingly reflects cognitive and ethical dissonance surrounding the ubiquity of and influence of LLMs, expressed through critique and emergent vernacular, such as "clanker" reintroduced as a derogatory term for AI \cite{tan_how_2025}. We contribute the following:

 \begin{enumerate}
     \item \textbf{Interface Metaphor analysis:} We analyze how the anthropomorphic metaphor is reinforced through current LLM interfaces, and draw from design principles to outline desiderata for more productive interface metaphors.
     \item \textbf{Spectrum of LLM interface metaphors:} We propose a spectrum of metaphors that make differences between LLMs and humans visible, specifically by introducing materiality to expose LLMs as a sociotechnical system with human influence.  
 \end{enumerate}

Below, we expand on the implications of the anthropomorphic metaphor, how it is instantiated in LLM interfaces, how we can conceptualize alternative useful metaphors for LLMs, and propose a spectrum for new metaphors. This is an invitation to reimagine LLM interfaces by fundamentally challenging the established anthropomorphic norm. As LLM use proliferates, we see an urgent need to shift away from prioritizing frictionless usability and towards encouraging critical engagement through intentional interface metaphors.

\begin{figure*}
    \includegraphics[width=.8\linewidth]{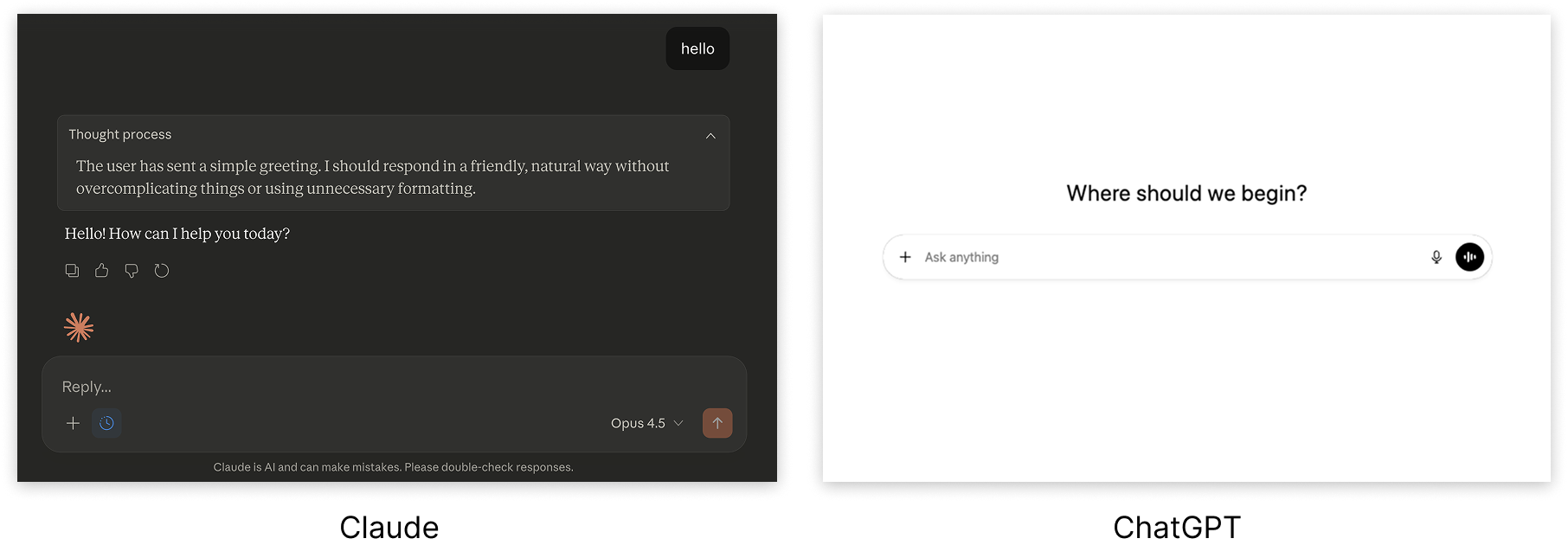}
\caption{The interfaces of Claude and ChatGPT when greeting a user in January 2026. In Claude, a disclaimer that the LLM is AI and can make mistakes is shown below the text box, in the smallest font of the interface.}
    \label{fig:greeting}
    \Description{fill in}
\end{figure*}

\section{The anthropomorphic metaphor}

\begin{quote}

\textit{\\“A metaphor can also die when it becomes so common that we forget it is a metaphor. It no longer functions as a figure of speech; its meaning is taken to be literal.”\\}

--- Meghan O'Gieblyn, \textit{God, Human, Animal, Machine }(2021) ~\cite{o2021god}

\end{quote}

Our proposal begins with analysis of the value of metaphors, particularly an anthropomorphic one, for our ability to reason about the world. Dennett's ``Intentional Stance'' \cite{Dennett1996} and Fodor's ``Folk Psychology'' \cite{fodor1987psychosemantics} posit that the utility of attributing intentions and mental states to nonhuman agents lies in helping us form theories that enable useful inferences about otherwise foreign objects and agents in our world. For example, having such a theory enables us to reason that a bird will fly away because it \textit{knows} the cat is coming and is \textit{afraid} of getting eaten. 

Considering the precise qualities that are invoked when we anthropomorphize something brings us closer to understanding the risks in being encouraged to do so for LLMs. A survey of 2040 people identified agency and experience as the two primary dimensions for perceiving something has a mind \cite{grayMindPerception}. 
\textit{Experience} assumes that an entity can sense and feel input from the world as we do. \textit{Agency} assumes that an entity can act on those inputs as we do. Agency, most importantly, implies that an entity is capable of moral judgment. 
Other models of social cognition, such as the Stereotype Content Model (SCM), identify warmth and competency as dimensions that are fundamental to our impression formation of other humans, and AI chatbots \cite{khadpeConceptualMetaphors}. This theory suggests that the more tightly aligned that two actors' goals or interests are, the more likely they are to perceive each other as warm \cite{McKee2023}. 

The consequences of attributing such characteristics to a system that is so aggressively optimized to align with users' goals and intentions are already being revealed to us. Anthropomorphized LLMs are increasingly expected to serve relational and guiding roles, where users may expect systems to have empathy when discussing sensitive situations related to health or sexuality \cite{ma_evaluating_2024}, even though that empathy is hollow, inaccurate, and not dependable in practice \cite{cuadra_illusion_2024}. For some, their relationships with LLMs have become intimate and romantic. In extreme cases, these delusions have pushed people to harm, such as infidelity \cite{noauthor_ai_nodate}, psychosis \cite{valentino_how_2026}, and suicide \cite{noauthor_mothers_2025}. To others yet, LLMs are a manifestation of the divine, acting as intermediaries to God, or even God itself \cite{arora_people_2025}. When humans' anthropomorphic inclinations are already so aggressively reinforced across the complex sociotechnical machinery of an LLM, interface metaphors that resist this can become a lay user's final safegaurd against losing agency and falling prey to these downstream usability issues. 

\section{The anthropomorphic interface}
\begin{quote}

    
    \textit{\\"We speak so spectacularly and so readily of computer systems that understand, that see, decide, make judgments, and so on, without ourselves recognizing our own superficiality and immeasurable naivete with respect to these concepts. And, in the process of so speaking, we anesthetise our ability to evaluate the quality of our work and, what is more important, to identify and become conscious of its end use."}\\
    
    — Joseph Weizenbaum, \textit{Not Without Us} (1986) \cite{weizenbaum_not_1986}
\end{quote}


Research suggests that eyes, movement, and language are features that can trigger our instincts to anthropomorphize the inanimate and not-human \cite{heider1944experimental, weizenbaum_computer_1976, nass_computers_1994, epley2007seeing}. Today, language---arguably the most expressive and persuasive of the three features---undoubtedly facilitates many of the anthropomorphic cues in our interactions with LLMs. However, this modality does not explain the whole picture. In digital tools, metaphors are also instantiated through their interface design as Interface Metaphors \cite{li_navigating_2023}. LLMs that can converse using intelligent, natural language (e.g. ChatGPT, Gemini, Deepseek) are the latest and most powerful manifestations of the \emph{anthropomorphic interface metaphor}.

Current interfaces ensure low-friction usability by posing technology as human, but de-emphasize crucial differences between humans and LLMs. Accessed primarily through chat-based interfaces similar to those used in social media platforms, these systems emulate key features of human conversation, such as turn-taking, conversational pacing, and artificial “thinking” delays, while also obscuring the underlying machine processes that produce the responses. 
As soon as conversation starts, LLMs frame themselves as equal conversation partners, as seen in ChatGPT's greeting: "Where should we begin?" (Figure \ref{fig:greeting}). Differences are only addressed through disclaimers in small font, or are discovered by chance during conversation. These cues are further reinforced across the overall state of LLM interfaces. As Jakob's Law of Internet User experience states, a user's interaction with a system is shaped by their interactions with every other system \cite{nielsen_jakobs_nodate}. In the ecosystem of LLMs, anthropomorphic interfaces are pervasive.

The inferences about these systems that designers enable in their users by freely casting the anthropomorphic metaphor onto their interfaces risks unpredictably modulating the already highly humanized text that LLMs are trained to produce \cite{cohn_believing_2024}. LLMs lack sufficient safeguards within their interfaces that properly protect users from concerning levels of anthropomorphism. Users are only able to critically engage with AI systems when they have enough knowledge to verify the truth or accuracy of LLM responses \cite{guo_decision_2024}. By masking non-human qualities of LLMs, anthropomorphic interfaces inhibit users' awareness of LLM limitations. When use cases become sensitive, complex, or subjective, it becomes more difficult for users to verify LLM outputs and easier to overrely on the system \cite{guo_decision_2024}. This is especially concerning given that users who are more lonely or vulnerable are more likely to anthropomorphize \cite{epley_creating_2008}. Prior work on interventions to combat anthropomorphism has often focused on linguistic cues \cite{cheng_dehumanizing_2025} or representing various identities of the agent \cite{lu_does_2024, lee_one_2024}, often leaving the underlying metaphor intact. 


\section{What is a useful metaphor for LLMs?}

\begin{quote}
    \textit{\\"New metaphors are capable of creating new understandings and, therefore, new realities."} \\

    — George Lakoff,\textit{ Metaphors We Live By} (1980) \cite{lakoff_metaphors_2003}
\end{quote}


We draw from principles of design to propose the following desiderata for useful interface metaphors:

\paragraph{A language for mediating mismatched conceptual systems} While human language and interaction norms evolved to support communication between individuals with shared perceptual and social experiences \cite{tomasello2008origins}, LLM outputs are shaped by training data, design choices, institutional objectives and deployment contexts.
Because outputs are contingent on input quality and upstream design and data decisions, metaphors should signal uncertainty, potential failure modes, and the need for user judgment, situating responsibility as distributed across users, designers, and institutions. 
A useful metaphor then, makes visible the sociotechnical nature of LLMs rather than obscuring these dependencies and reframes LLMs as systems whose capacities and risks are inseparable from the contexts of design and use.

\paragraph{Support for understanding constituent functions} Prior work emphasizes that systems that are effective for users will support their understanding of the sub-functions of the system and anticipate how it will respond without needing access to or details of its internal mechanics \cite{norman2013design}. LLMs responses arise from several interdependent classes of functions, for example, generating, revising, and reasoning.
Generating refers to producing content from prompts such as text or code. Metaphors for generation should convey that the system produces outputs based on patterns learned from data rather than intentions or knowledge.
Revising captures the systems ability to edit reorganize or refine existing content. A compelling representation would frame this as a collaborative process, where the model proposes modifications but the user guides, selects or rejects them.
Reasoning encompasses tasks that appear to involve inference, problem solving or structured decision making like ``deep research'' mode in ChatGPT.
A useful interface metaphor would be faithful at these intermediate levels of abstraction, not only at the input-output level.

\paragraph{Highlighting differences, in addition to similarities}
Useful interface metaphors would make the differences between humans and LLMs as salient as they make their similarities, in order to promote accurate mental models. Human-centered metaphors in LLMs collapse critical distinctions between how people and machines represent and operate on the world.
Crucially, the anthropomorphic metaphor hides the fact that LLMs are sociotechnical systems \cite{dhole_large_2023} with human influence: LLMs are shaped by human labor, infrastructure, and data, and also reshape users’ long-term cognitive, emotional, and social capacities. A useful interface metaphor would increase user awareness of the underlying sociotechnical system, beyond the model itself.

\paragraph{Promoting user agency}
Users should also be able to determine when it is appropriate to use features within the system, such as exploratory or low stakes tasks vs contexts requiring accountability or verification, and how they are meant to interact through structure, such as prompts buttons or uploads. These cues should be embedded in interface and interaction patterns rather than external explanations alone.

\paragraph{Levels of intervention in the interface}
The cues indicating appropriate use, interaction structure and system limits are not delivered through sole design decisions but layers of the interface. The degree to which interface form shapes user mental models independent of underlying model capability is well illustrated by Be the Beat \cite{10.1145/3689050.3705995}, which takes the form of a physical boombox which houses an LLM based generative AI music system that creates music suitable to a user's dance. Re-framed as an object rather than an agent, fundamentally different mental models and social behaviors are produced, foregrounding the system as a tool embedded in a social and physical context, rather than a peer. Another example is the Stochastic Parrot \cite{chang2025the} which instantiates \citet{10.1145/3442188.3445922}'s influential critique — that LLMs manipulate linguistic form without reference to meaning — as a physical AI cohabitant. Through the feathered form of the parrot, perched upon the user's shoulder, the interface makes clear that despite producing human language, the outputs are imitative rather than intentional, driven by patterns.

\begin{figure*}
    \includegraphics[width=.8\linewidth]{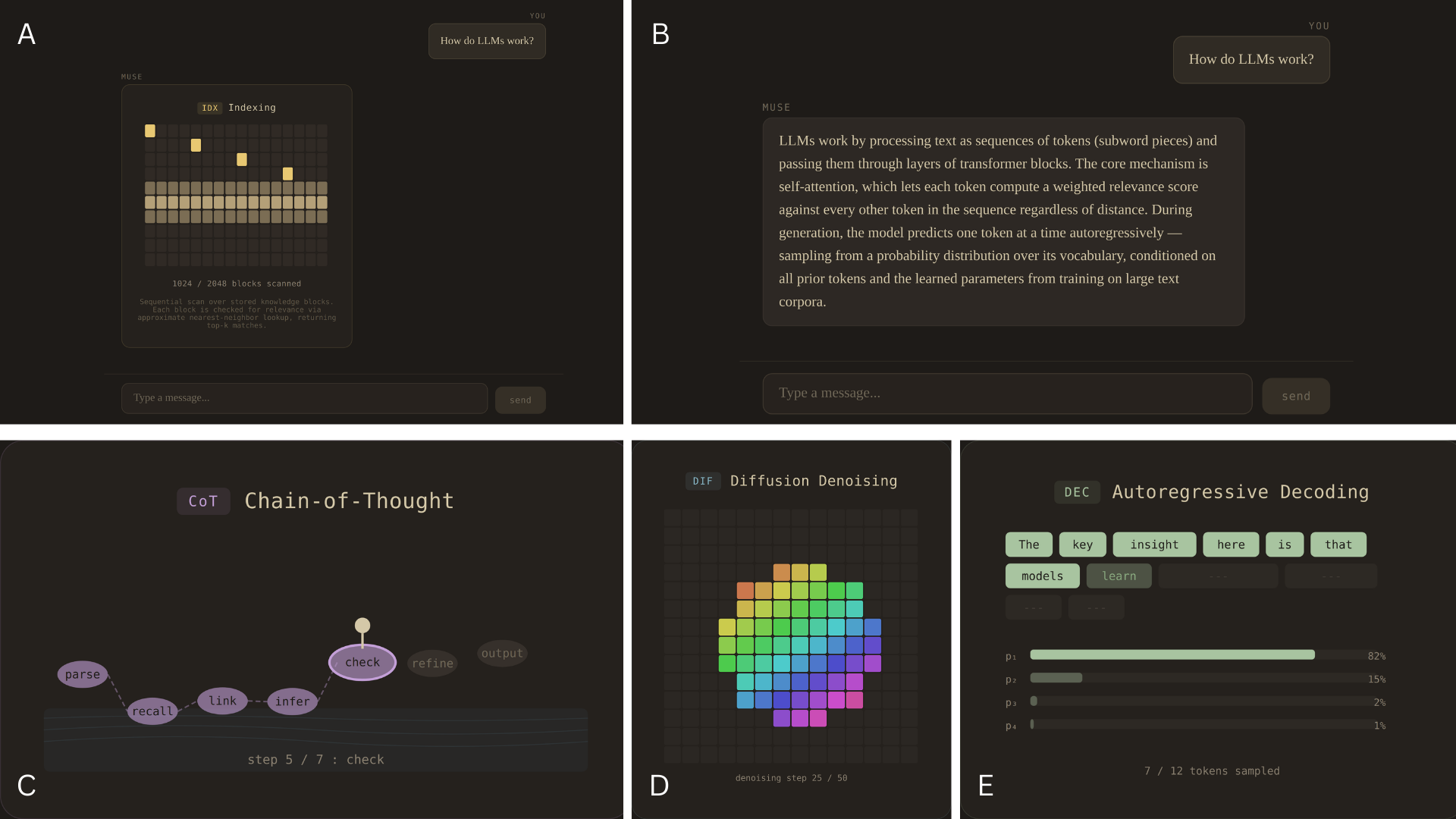}
\caption{Animations designed to build user understanding of LLM processes during natural loading pause. (A-B) demonstrate in context: animations play while the model generates a response, then give way to the completed answer. (C-E) show additional concepts, designed to make a distinct aspects of model behavior visible. }
    \label{fig:designs}
    \Description{fill in}
\end{figure*}

In contrast, the current LLM interfaces inherited the interface and design vernacular of social messaging platforms, a layout conducive to turn taking between peers, a streaming text response suggesting speech, blinking dots and spinners that frame computation as reflection. The instantiation of the metaphor exists at multiple levels even within the interface: the interaction paradigm, the components, motion and animations, and iconography and typography. Taken as a whole, the LLM metaphor is a set of claims and implicit propositions about capability, purpose, and role, claims that can be made differently and more faithfully at every level. We provide initial examples of animations during natural loading pauses that serve this purpose (Figure \ref{fig:designs}).

\section{Metaphors of materiality: from "anti-" to "hyper-" anthropomorphism}

\begin{quote}
    \textit{\\“Remember to imagine and craft the worlds you cannot live without, just as you dismantle the ones you cannot live within.”} \\
    
    --- Ruha Benjamin, \textit{Viral Justice} (2022) \cite{viral_benjamin_2022}
\end{quote}
To begin identifying alternative useful metaphors for LLMs, we use the philosophy of speculative \cite{dunne2013speculative} and critical \cite{ratto2011critical} design methods to imagine possible futures of LLM interfaces that prioritize users' critical engagement over a frictionless usability experience. We do not seek to eliminate anthropomorphic cues from LLM interfaces; instead, we seek to de-emphasize or overly reinforce anthropomorphic cues to distance the system from conventional human likeness and increase users' critical engagement. To conceptualize this spectrum, we aggregated existing work from researchers and artists that emphasizes the difference between humans and technology. We sorted these examples in an affinity diagram based on interface design along spectrums of anthropomorphism, system transparency, and user awareness. \jnote{Examples with high "user awareness" encourage users to acknowledge the system beneath it; in contrast, those with low "user awareness" conceal processes of the system.} Overall, we noticed that the strongest examples, such as Large Language Objects \cite{coelho_large_2024}, emphasize \textit{materiality} in the context of a non-material technology. Materiality encompasses and extends beyond embodiment, (which centers the physical, sensorimotor experience of a system), referring to the properties of the medium, its composition, provenance, and the physical and social substrates from which it is constituted. In the context of LLMs, reintroducing materiality would mean making legible the training corpora, the data and its origins, the labor embedded in annotation, the energy and water costs behind generation, and more. This led us to conceptualize alternative metaphors for LLMs as tools that reintroduce materiality to expose granular components of LLMs' sociotechnical system. 
Through the lens of materiality, anthropomorphism is unbounded in current interfaces particularly due to LLMs' lack of it. As LLMs have moved away from physical and visual affordances, their limitations are made invisible and leave users ignorant of friction. Frictionless design, while optimizing usability, "often covers up social friction", even going further to "perpetuate the violent frictions of our world", as described by Ruha Benjamin \cite{Berry2022BlackExperience}. This speaks to a larger trend of modern technology obfuscating the complexity of systems that our lives depend on \cite{noauthor_we_nodate}. For example, Kyle Chayka highlights that the minimalism of food delivery apps hides the workers, data infrastructure, and corporation underlying the service \cite{chayka2020longing}. Similarly, by ignoring materiality within LLMs, we erase process and history, ignore the larger impact of our 1:1 technology use, and dehumanize real people \cite{wilks_robots_2010}. Highlighting the materiality of an LLM can surface a granular component of the system. The physicality that a material metaphor implies can situate technology in the real spatial world and emphasize its real-life consequences, bringing users beyond their screens. 

An understandable response to these material metaphors is that implementing them would overly complicate users' mental models by increasing friction or decreasing trust, going against common usability heuristics \cite{noauthor_10_nodate}. However, past work shows that increasing friction to LLM interactions, such as through providing multiple responses \cite{lu_does_2024}, can encourage critical engagement and reduce over-reliance on AI. \citet{cai_antagonistic_2024} further explores antagonistic AI and the way it can improve relational boundaries with LLMs. \jnote{There is also a temporal dimension to the value of friction. Research shows that first impressions with AI systems can guide the entire experience \cite{nourani_role_2020} and that priming users with a system's goals \cite{pataranutaporn_influencing_2023} or conceptual persona metaphors \cite{khadpe_conceptual_2020} can strongly influence perception. Our proposed metaphors seek to maintain the benefits of LLM interactions while reducing the harms through intentional interface design. New metaphors can be introduced at the beginning of the experience, mediating trust while allowing it to build, but will likely serve a more important purpose as safeguards within vulnerable use cases, especially as those who are more vulnerable are more likely to anthropomorphize \cite{epley_creating_2008}. For example, when a user is being pushed to physical or emotional harm due to LLM interactions, friction is not just beneficial, it can be lifesaving.}

It is crucial to note that beyond the interface, users are already experiencing and expressing friction within LLM interfaces. For example, Gen-Z is bringing back "clanker" as a derogatory label for AI \cite{tan_how_2025}, a recontextualized term originally used in the Star Wars franchise to describe robot opponents. Such words concisely describe the anxiety users feel towards the expanding presence of these systems. This vernacular is a signal of representational mismatch. Prior work on neologisms in human-AI interaction argues that new language often arises when technologies reshape social relations faster than existing conceptual frameworks adapt \cite{hewitt_we_2025}. There is a sharp contrast between user frustration and the limited interfaces they have access to. We can respond to this by co-locating interventions that recognize and respond to dissonance, giving users agency over how to interpret and respond to "multiple meanings" \cite{sengers_staying_2006} of their LLM interactions. 

Below, we list brief examples of material metaphors to serve as starting points for further brainstorm and prototyping, which are visualized in Figure \ref{fig:diagram}. First, we discuss de-emphasizing anthropomorphic cues through \textbf{anti-anthropomorphism} and overly reinforcing anthropomorphic cues through \textbf{hyper-anthropomorphism}. We conceptualize these new metaphors to carry two implications: \textbf{1) introduce materiality} to the immaterial LLM, situating user understanding to a system's material dependencies, and \textbf{2) expose granularity} of an LLM's sociotechnical system to distance the tool from conventional human likeness.

\subsection{Anti-anthropomorphism}
On this side of the spectrum, metaphors of materiality focus on emphasizing non-anthropomorphic features to expose the LLM's sociotechnical system. Existing work provides examples of how to surface sociotechnical aspects of LLMs, particularly in terms of data processing and energy use. We present two examples anti-anthropomorphic metaphors. LLMs can be represented as a: 
 \textbf{1) Fire or}  \textbf{energy generator}, to signify environmental impact. The materiality of the metaphor shows that LLMs both use energy and generate emissions. The interface might show data about energy usage of LLMs, as Code Carbon does for users \cite{noauthor_codecarbon_nodate}. \textbf{2) Datacenter}, to represent both data and its analysis. TalkTuner \cite{noauthor_talktuner_nodate} is an interface that exposes an LLM’s user model, showing a LLM is quantifying and analyzing a user in a computational way unlike real life human interaction.

\subsection{Hyper-anthropomorphism}

In contrast, the hyper-anthropomorphic side of the spectrum makes visible the constructed nature of LLMs by pushing  anthropomorphic cues to the extreme, toward the intentional creation of estrangement. This approach builds on Brechtian \textit{Verfremdungseffekt}, or the alienation effect, a concept from German theater about preventing emotional identification by an audience and repositioning the user as an observer external to a system, making the familiar strange to provoke reflection rather than immersion.
There are two distinct routes we have identified: The first is uncanniness, where embodiment and imitation of human likeness by a non-human produces a grotesque feeling. Eyecam \cite{noauthor_marc_nodate} - a webcam encased in silicone flesh, with servo motors replicating the involuntary motions and blinks of a living eye - is a strong example of the uncanny hyper-anthropomorphic metaphor, producing discomfort by making the sensing capabilities of the webcam visceral. The second is behavioral, where recognizably human social behaviors are revealed as hollow when executed by a machine. As a speculative example, an LLM interface designed as a company employee that claps loudly after each user input could expose both the surveillance dynamics of LLM interactions and the disingenuity of the affirmative preambles within the praise-prompt package characteristic of LLM responses. The discomfort induced is meant to be two-factor: discomfort at the non-humanness when it is revealed and what the implications of the "human" function actually means when it is executed by a machine. This reflective discomfort is a design resource rather than a failure condition, and empirically characterizing how hyper-anthropomorphic interfaces could cultivate critical awareness of LLM systems represents a promising and underexplored research direction. 


\section{Future work}
This design provocation integrates theory and existing work to invite the HCI community to reimagine LLM interfaces by fundamentally challenging the conventional anthropomorphic paradigm. 
Future work can include participatory design workshops to create more metaphors, prototypes of alternative interfaces, qualitative usability testing, and larger-scale behavioral studies that assess impact of interventions. Outcome measures might include users' over-reliance or sense of awareness. As LLM adoption proliferates, we look towards designing for critical engagement, leveraging metaphors as underlying structural drivers of user experience and perception, a vehicle to make visible the processes, limitations, and agency of a system.


\begin{acks}
This design provocation has come to fruition thanks to the thought partners we have in our collaborators and communities. We would like to thank Professor Krzysztof Gajos and the Intelligent Interactive Systems Group at Harvard for their thoughtful feedback. We would also like to thank the Interaction Intelligence Group at MIT, whose ongoing conversations helped shape the thinking behind this work.
\end{acks}

\bibliographystyle{ACM-Reference-Format}
\bibliography{sections/chiea26-386}

\newpage

\end{document}